# In-situ analysis of small microplastics in coastal surface water samples of the subtropical island of Okinawa, Japan

Christina Ripken[a,b], Domna G. Kotsifaki[a] and Síle Nic Chormaic[a]

[a]Light-Matter Interactions for Quantum Technologies Unit, Okinawa institute of Science and Technology, 1919-1 Tancha, 904-0495 Onna, Okinawa, Japan

[b]Marine Genomics Unit, Okinawa institute of Science and Technology, 1919-1 Tancha, 904-0495 Onna, Okinawa, Japan

## ARTICLE INFO

## ABSTRACT

Marine plastic debris is widely recognized as a global environmental issue. Sub-micron plastic particles, with an upper size limit of 20 $\mu m$, have been identified as having the highest potential for causing damage to marine ecosystems. Having accurate methods for quantifying the abundance of such particles in a natural environment is essential for defining the extent of the problem they pose. Using an optical micro-Raman tweezers setup, we have identified the composition of particles trapped in marine aggregates collected from the coastal surface waters around the subtropical island of Okinawa. Chemical composition analysis at the single-particle level indicates dominance by low- density polyethylene, which accounted for 75% of the total sub-micron plastics analysed. Our results show the occurrence of plastics at all test sites, with the highest concentration in areas with high human activities. The average, smallest sub-micron plastics size is $(2.53 \pm 0.85)$ $\mu m$ for polystyrene. We also observed additional Raman peaks on the plastics spectrum with decreasing debris size which could be related to structural modification due to weathering or embedding in organic matter. By single-particle level sub-micron plastics identification, we can begin to understand their dispersion in the ocean, and define their toxicity and impacts on marine biodiversity and the food chain.

## 1. Introduction

Plastic polymers are a versatile, widely used material fully integrated in our daily lives. In the environment, plastics accumulate because of their recalcitrant nature [1]. Once plastic items are discarded in the environment, they often end up in waterways and are ultimately transported to the ocean [2]. The first report on the emergence of small plastic particles in the oceans drew worldwide concern [3]. Because most plastics undergo very slow chemical or biological degradation in the environment, the debris can remain in the ocean for years, decades, or even longer [4]. Moreover, plastic debris can entrap marine fauna [5] and be ingested by a wide variety of animals, ranging in size from plankton to marine mammals [6–8]. Ingestion of marine plastic fragments and fibres into the trophic chain may cause human health problems. Furthermore, field observations and oceanographic models show that five subtropical ocean gyres are hotspots for plastic debris accumulation [9]. Supporting this, it has been reported that the global microplastic distribution across the oceans is estimated to be 236 thousand metric tons [10]. However, a discrepancy of orders of magnitude exists between these observations and the expected mass of microplastic in oceans, suggesting complicated export dynamics are at play.

While mesoplastic (5 mm – 2.5 cm) and macroplastic (> 2.5 cm) marine pollution has been extensively studied for many oceanic regions and across different ecosystems [11–16], sub-micron plastic pollution has been less of a focus [1, 6]. As proposed by the National Oceanic and Atmospheric Administration (NOAA), the term microplastics refers to very small, ubiquitous plastic particles < 5 mm in diameter [17]. They have been separated into two fractions, large (1 – 5 mm) or small (20 $\mu m$ – 1 mm) microplastics and the sub-micron fraction (20 $\mu m$ – 1 $\mu m$) plastics [18]. Like large microplastics, sub-micron plastics can adsorb and carry hydrophobic chemicals that have a potential biological and toxicological impact on the environment [19]. Therefore, a clear understanding of the interaction of sub-micron plastics with the environment, especially with living organisms, is essential to assess possible health hazards.

Currently, there is a need for reliable and precise identification of plastics without separating them from the matrices in which they were collected. The current protocols for quantification and characterisation of environmental plastic contamination is hampered by a lack of sensitive yet high-throughput methods. Commonly applied techniques for the analysis of plastics include a visual inspection or stiffness test [20], spectroscopy [21–23], transmission or scanning electron microscopy [24], and fluorescence imaging [7]. However, the chemical characterisation of single, plastic particles in a liquid environment is still limited. Therefore, techniques with selectivity and precision that enable the analysis of single particles *in situ* and in real-time are necessary.

Since the first demonstration of single microparticle optical trapping in 1986 [25], optical tweezers have emerged as a powerful tool for controlling particles in fluids [25]. Optical tweezers use a highly focussed laser beam to trap and manipulate (typically) dielectric particles from 10 nm to 100 $\mu m$ in diameter. The particle is trapped near the focal spot due to scattering and gradient optical forces [26, 27]; the scattering force is from radiation pressure of the light beam along its direction of propagation and the gradient force pulls the particle towards the high-intensity focal point. The total optical force exerted on a particle is in the range of 100 fN to 100 pN depending on the difference between the refractive indices of the particle and the liquid medium, and the intensity of the laser beam. The ability to measure such small forces has opened the way for many new experiments in physics, chemistry, biophysics, and nanotechnology. Important examples include the development of multiple optical trapping [28], a variety of biophysics measurements on single biomolecules [29], cell-sorting applications [30], optical binding of particles [31], and trapping in subwavelength fields created by plasmonic nanostructures [27].

*Corresponding author: Domna G. Kotsifaki: domna.kotsifaki@oist.jp

The combination of optical tweezers with a range of different optical read-out techniques has enabled various types of single-particle investigations to be done. With regard to *in situ* sub-micron plastic analysis, optical tweezers micro-Raman spectroscopy (OTRS) is an ideal option. Raman spectroscopy has been used to analyse single cells and biomolecules suspended in an aqueous environment. The first combination of micro-Raman spectroscopy and optical tweezers was presented in 1994, when Urlaub et al. investigated a polymerisation reaction in optically trapped emulsion particles [32], while Raman tweezers entered the field of biophysics in 2002 with studies of single cells and organelles [33–35]. Recently, the OTRS technique has been used for chemical qualitative analysis of a variety of plastic particles with sizes in the sub-20 $\mu$m regime in a seawater environment [36]. The authors successfully discriminated between plastics and mineral sediments at the single-particle level, overcoming the limitations of conventional Raman spectroscopy in a liquid environment [36].

Compared to the rest of the world, there is limited information on plastic pollution of seawater in a "blue zone" area, regions of the world in which exceptional longevity has been recorded [37]. In this work, we have collected environmental samples around the subtropical island of Okinawa and analysed them using an optical tweezers micro-Raman spectroscopy technique, to determine the occurrence, the average size and the polymer type of sub-micron plastic particles. Therefore, we improve our knowledge of the extent and magnitude of sub-micron plastics pollution in the ocean around a blue zone region. This gives an overview of the current state of pollution, as well as how this pollution correlates with population and industrial densities on the Okinawa island.

## 2. Materials and Methods

The main island of Okinawa (26.2124° N, 127.6809° E) is part of the Ryukyu Island Arc (inset in Figure 1) and consists of uplifted coral reefs and, particularly in the northern half, igneous rock (solidified magma or lava). It is surrounded by fringing reefs, making the water intake that reaches the beaches more reliant on surface waves and wind [38]. Land-based pollution originating on Okinawa is more likely to be found in the bigger bays of the island. The six sampling sites were chosen to provide a road map of the pollution distribution of Okinawa (Figure 1). Sites differ in population density and industry in and around the respective bay re-gion. Additionally, Okinawa has been previously deemed as a "blue zone" [39]. Therefore, it is crucial to monitor ocean pollution as it may adversely affect the residents' longevity in such a region.

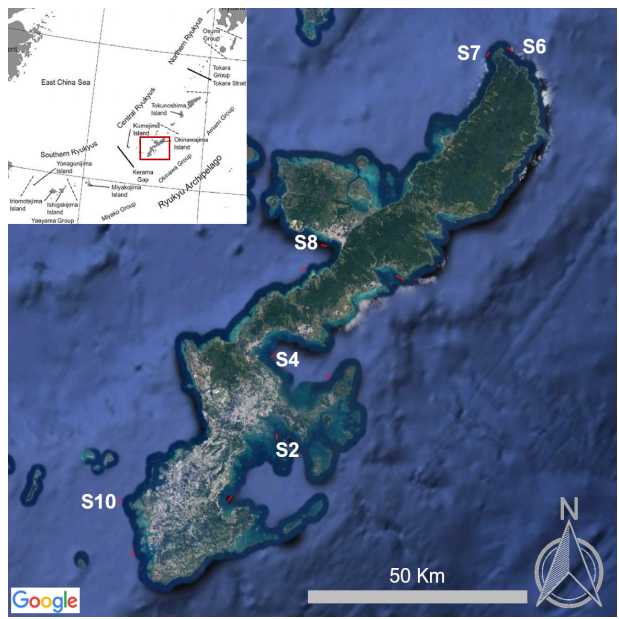

**Figure 1:** Map of field study area in Okinawa with the locations of the six towing stations from which particles were collected and analysed [40]. Inset: Geographical location of Okinawa in the Ryukyu Island Ark [41].

### 2.1. Study Region

To quantify sub-micron plastic abundance in the surface waters around Okinawa, the water samples were collected over 24 hours in September 2018 with the Okinawa Prefectural Fisheries (OPF) and Ocean Research Center (ORC) ship, Tonan Maru. The cruise was designed to obtain an overview of the sub-micron plastic pollution around Okinawa.

### 2.2. Field Sampling

The sampling was performed using a manta trawl (Hydro-Bios Manta, 300 $\mu$m net, net opening of 15 cm× 30 cm, with a Hydro-Bios Mechanical Flow Meter No.: 438 110). The manta net was deployed off the starboard side of the boat to the surface and trawled for 15 min at 2 - 3 knots covering an average distance of 1 km and filtering an average volume of 856.8 L. After trawling, the nets were washed down twice before transferring the contents into 450 mL glasses. All samples were stored in a climatised environment until processing the next day. Between each sample, the Manta net was backwashed with seawater and the collector at the end of the net was washed separately to limit cross-contamination between sampling locations.

### 2.3. Laboratory Analysis

Each of the manta trawl samples was filtered over a 300 $\mu$m sieve to remove large organic materials and microplastics.

The remaining samples, with particles smaller than 300 $\mu$m, were used for further analysis. To limit contamination of the samples via clothing or air, all samples were filtered, sorted, and prepared onto the microscope slides in a positive pressure chamber. Only cotton clothing was worn during sampling and preparation.

## 2.4. Optical micro-Raman Tweezers Spectroscopy

The OTRS system we used consists of a Nd:YAG laser beam ($\lambda = 532$ nm with 17 mW of power at the sample plane) focused using a high numerical aperture (NA = 1.3) oil immersion objective lens (Plan-Neofluar 100×, Carl Zeiss) onto the seawater sample, as shown in Figure 2. The trapping laser beam is integrated into a Raman spectrometer (3D Laser Raman Microspectrometer Nanofinder 30). The high NA of the lens ensures trapping of the sub-micron particle and provides the necessary laser intensity needed to maximise its Raman signal. Using adhesive microscope spacers, a microwell was formed on the microscope glass slide, and trapping occurs in the 20 $\mu$L sample solution under a glass cover slide. The microwell contains sub-micron plastics and nanoparticles in seawater from the sampled regions around Okinawa. As a control, a microwell with 10 $\mu$l milli-Q water aqua was prepared on the same microscope slide. The microscope slide was mounted and fixed on top of a translation stage. Using this setup, the OTRS technique allows us to identify small microplastic fragments trapped by the laser.

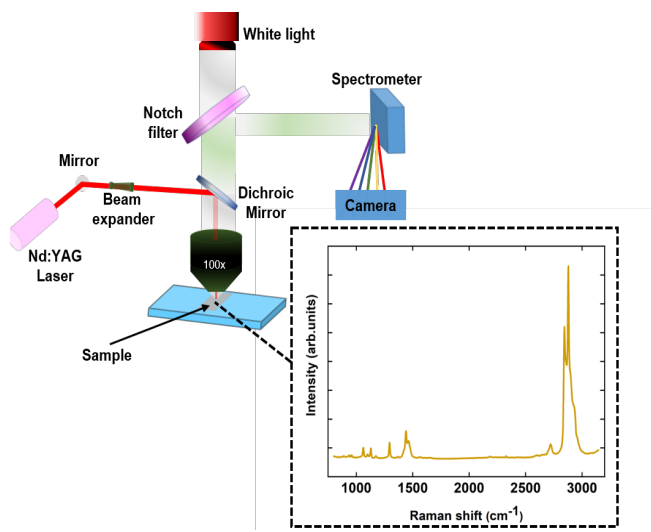

**Figure 2:** A schematic illustration of the micro-Raman optical tweezers setup used in our experiments. Inset: characteristic spectrum of a polyethylene (PE) particle of 15 $\mu$m diameter in a seawater environment.

## 3. Results and Discussion

Ocean surface water samples from several bays around Okinawa have been collected (see Figure 1) and analysed, as shown in Figure 2. The chemical identification of optically trapped particles within the seawater sample was accomplished by employing OTRS after background subtraction. Notably, the majority of the sub-micron plastics are quasispherical. A shift of the Raman peaks or alteration of the bands can be expected due to their crystalline structure and level of degradation. Table 1 lists the most dominant types of plastics which we have identified in seawater around Okinawa. Moreover, we include their hazard score in order to indicate their potential impact on the environment and human health.

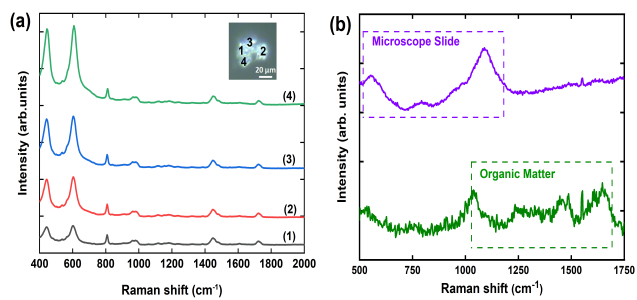

**Figure 3:** (a) Raman spectra of a polyvinyl chloride (PVC) quasispherical microplastic of 47.8 $\mu$m diameter found at station S10, near Naha. Each spectrum relates to a different region on the trapped particle. Inset: microscope image of the plastic in which different spectral regions are labelled. The intensity of the Raman signal may indicate the difference in weathering. (b) Additional Raman spectra peaks that typically appear in the signals from the trapped particles in our samples. Purple (upper) curve: spectrum from the microscope slide. It displays the glass solid-state structure with no long range transnational symmetry manifesting – the peaks are very broad with widths up to several hundred wavenumbers. Green (lower) curve: spectrum from organic matter found in the samples with CCO stretching (around 1000 cm$^{-1}$) and CH$_3$ and CH$_2$ deformations (1250 cm$^{-1}$ to 1750 cm$^{-1}$) in the Raman spectrum. The individual peaks vary depending on the organic matter of the trapped particles.

Figure 3(a) shows Raman spectra for a quasispherical, optically trapped microplastic diluted in seawater from a sample collected near the Naha region (S10 in Figure 1). Naha is the capital of the Okinawa Prefecture and the sample location was next to the industrial port and commercial airport. Despite being a heavily commercialised area, Naha has an estimated population of 318,270 inhabitants, representing almost 30% of the total population of Okinawa island and a population density of 8,043 people/km$^2$. In Figure 3(a), we have identified the characteristic Raman peaks of polyvinyl chloride (PVC) (Table 2-left). Additionally, we investigated the Raman signal of the optically trapped microplastic for various positions on its surface. We note that the intensity of the Raman signal changes. This could be due to the difference in weathering of the material at various places. In total, 51 particles were analysed from station S10, with approximately 19.6% being plastic such as polyethylene (PE) or polyvinyl chloride (PVC). This is the second highest percentage of plastics that we have found in seawater around Okinawa. This result correlates well with a recent study by Kitahara et al. [44], finding small plastics in road dust on Okinawa. Although the population density is highest in Naha and most land use is urban [45, 46], plastic particles in the road dust were lower in front of our station S10 [44]. Another reason for finding the second highest plastic pollution at this station is the location outside of a bay. Although many rivers discharge on this side of Naha city [47], the pollution will not be trapped on this side of the island.

**Table 1**
**Polymer types found as nanoplastics around Okinawa**: Detailed information for polymers identified in this study, including monomer, density, hazard score, and life span [42, 43].

| Polymer | Abb | Monomer | Density (g/cm$^3$) | Hazard score | % of total particles analysed | Life span (years) |
|---|---|---|---|---|---|---|
| Polyethylene | PE | Ethylene | 0.91-0.96 | 11 | 10.94 | 20 |
| Polypropylene | PP | Propylene | 0.85-0.94 | 1 | 0.61 | >100 |
| Polyvinyl chloride | PVC | Vinyl chloride | 1.41 | 10551 | 0.61 | 140 |
| Polyamide (Nylon) | PA | Adipic acid | 1.14-1.15 | 47 | 1.52 | >20 |
| Polystyrene | PS | Styrene | 1.05 | 30 | 0.91 | 50 |

**Table 2**
**Raman peaks of**: **(Left)**: Polyvinyl chloride (PCV) has a hazard score of 10551 [43], making it one of the most toxic plastics based on hazard classification of monomers. PVC is mostly used for cables, pipes & fittings, window frames, and flexible films for water proofing. **(Right)**: Polyethylene (PE) is the most common plastic used in daily life. Primarily used for packaging, resulting in air trapping items such as bottles and plastic bags. Combined with its low density this leads to a majority of the PE floating at the ocean's surface [36].

| PVC : $\nu$ (cm$^{-1}$) | Vibration |
|---|---|
| 1724 | Ester CO stretching |
| 1434 | $CH_2$ symmetric deformation |
| 1325 | $CH_2$ twisting |
| 610 | Crystalline C-Cl stretching |

| PE : $\nu$ (cm$^{-1}$) | Vibration |
|---|---|
| 1058 | CC symmetric stretching |
| 1123 | CC anti-sym stretching |
| 1286 | $CH_2$ twisting vibration |
| 1408 | $CH_2$ bending |
| 1429 | $CH_2$ symmetric deformation |
| 1450 | $CH_2$ scissor vibration |

Plasticisers, dyes, and weathering can change the Raman spectra, adding additional peaks to the spectra of the different polymers as well as changing relative intensities and accuracy. These additives are often harmful and can leach from the polymer matrix [48]. In addition, the particles are often embedded in organic material, which can also add peaks to the actual polymer spectra. In Figure 3(b), we show the Raman spectra of the microscope slide that we used in our experimental process and the organic matter found in the plastics. Based on these reference spectra we can distinguish the plastics from organic matter and identify their Raman peaks. Generally, we observed that most sub-micron plastics are embedded into organic matter (68.75%) while only 31.25% are free floating. No data is published on this ratio, as the methods for polymer identification used most often add a step of digesting the organic material in the sample first, to get a better Raman signal. This ratio is important to investigate further, as it can shed some light on the fate of the micro- and submicron plastic within the water column.

Figure 4(a) shows the Raman peaks of PE sub-micron plastics which were found in several areas around Okinawa, while in Table 2 (right) we present the modes attributed to

PE. We note that all the particles have the characteristic PE peaks spanning from 1000 cm$^{-1}$ to 1500 cm$^{-1}$ [36]. Figure 4(b) shows the Raman spectra for polystyrene (PS) and polypropylene (PP) particles which were not embedded in organic matter.

The most common plastic that was found in the seawater of Okinawa is PE with a percentage of 10.94% of the total particles analysed (see Table 1). The reason for the high percentage of PE could be its structural characteristics and lower density compared with the other polymer types found (see Table 1). It has more porous structures than other plastics and, as such, it may be more easily broken down into microscopic debris by sunlight, wind, and current erosion [49]. We notice surrounding organic matter overlays the PE Raman spectra in a microparticle of 5 $\mu$m diameter (red line in Figure 4(a)), which was collected from the Naha (S10) area. Additionally, an overlay of dyes or additives is observed in Raman spectra of sub-micron plastic with 5 $\mu$m diameter (purple line in Figure 4(a)) which was collect from the Kin (S4) area. The characteristic peaks of polystyrene (PS) (black line in Figure 4(b)) and polypropylene (PP) (blue line in Figure 4(b)) were identified at Nago (S8) and Nakagusku (S2) areas, respectively. PS is frequently found in the environment as a material from diverse uses such as packaging foams and disposable cups. Since it is mainly used for manufacturing of single-use products, a large portion of post-consumer production ends up into oceans [2], and re-mains there for several hundred years due to their resistance to degradation (Table 1). PP is used in the manufacturing of, for example, flip-top bottles, piping systems, and food containers, amongst others.

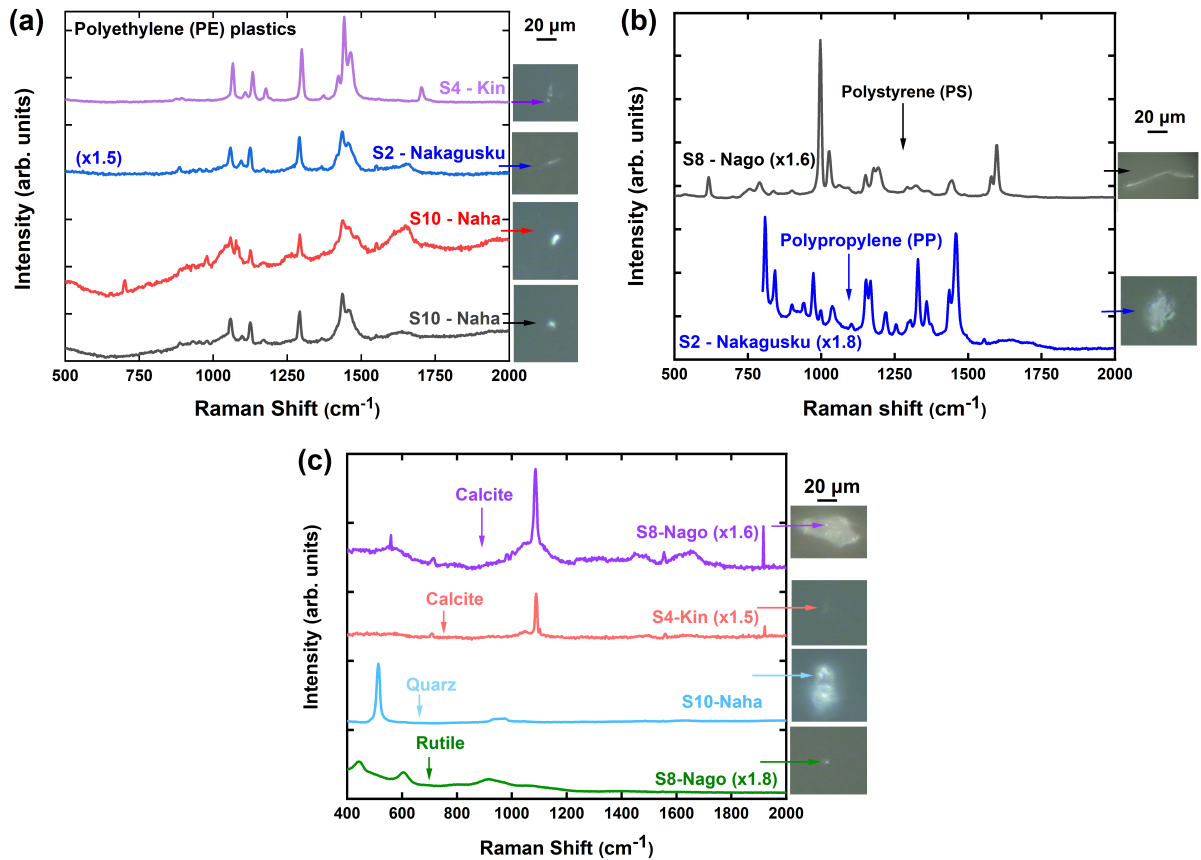

**Figure 4:** Raman spectra of optically trapped sub-micron plastics dispersed in seawater with their optical images: (a) Polyethylene (PE) spectra from different particles found in different locations. The red line (S10) has an organic matter overlay, while the purple (S4) and blue (S2) lines have additional peaks most likely from dyes or additives to the PE. (b) The black line indicates the Raman spectrum of polystyrene (PS) and the blue line of polypropylene (PP) found at Nago (S8) and Nakagusku (S2) areas, respectively.(c) Raman spectra of trapped sediment particles suspended in seawater with their optical images. In the purple line (S8-up) the calcite peaks are overlayed with organic matter, as observed on the optical image. The green line (S8-down) shows a rutile nanoparticle in which the Raman spectra is overlayed by the characteristic microscope slide Raman peaks.

Together with the plastics, trapped sediment micrometric and nanometric particles can also be detected (see Figure 4(c)). Specifically, we note that some particles have peaks at 512 cm$^{-1}$ and 472 cm$^{-1}$ [50], indicating trapped quartz particles (blue curve-S10 in Figure 4(c)). Particles of polymorphous CaCO$_3$ with one peak at 706 cm$^{-1}$ followed by a larger peak at 1088 cm$^{-1}$ indicate that they are most likely calcite and not aragonite or vaterite [51]. The origin of these particles is probably related to trace calcite-based contaminants. Finally, we find particles that display the spectral fingerprint of rutile (green curve-S8 in Figure 4(c)) with microscope slide signal overlay [52]. Rutile is a mineral composed primarily of titanium dioxide (TiO$_2$) and is the most common natural form of TiO$_2$. These sediment-derived particles are likely found because of high river input [47].

The abundance of sub-micron plastics in each station displayed a difference between the number of plastics collected in urban and less populated areas. The heterogeneity of the plastics at the sampling stations may be caused by several factors, predominately the closeness of point sources such as sewage outfalls, river outlets and run-off after heavy rain fall. Atmospheric input of micro- and submicron plastic from domestic activities such as traffic should also be taken into consideration. In Table 3, we summarise the plastic distribution around Okinawa. In total we have identified 282 particles by employing the OTRS method, of which 48 are sub-micron plastics. The plastic pollution observed for sub-micron plastics follows the population gradient of the island [45, 46]. There is a clear distinction between the northern (Figure 1), less populated part of Okinawa, and the southern part with high population density. While the south west side with the capital Naha has the highest population density, there is no bay on that side of the island. On the south east side of Okinawa, on the other hand, the big bay of Nakagusku (S2-Figure 1) is located. The cities of Nanjō (2.98% of population), Urasoe (7.90% of population), Ginowan (6.71% of population) and Okinawa city (9.74% of population) are located along that bay. While only a handful of rivers drain into this bay, these rivers have been found to have the highest levels of inorganic nutrients [47] on Okinawa. We analysed 44 particles from that bay and we calculated that 25.0% are plastics (8 of PE, 1 of PP, 1 of PS and 1 of PVC). We conclude that this is the area with the highest percentage of plastics due to the high population density (2,838 people/km$^2$). The southern part of the island has a high proportion of urban land use [46], which, in combination with high traffic density [44], leads to high anthropogenic pressure on the coastal ecosystem [53]. This results in a significant increase of sub-micron plastic percentage in the southern half of the island (t-test, two tailed $p = 0.0147$). This is in reasonable agreement with studies showing microplastic abundance in areas with an increase of intensive anthropogenic activities such as: urban areas with high population density [54], tourist beaches with high density of tourists [55], areas of intensive agriculture [56], as well as fishing and shipping activities [57].

**Table 3**
**Sub-micron plastic distribution around Okinawa**: Station split for population and industry around Okinawa following the respective gradients.

| Area name | Station names | Population distribution | Industry distribution | Volume checked [$\mu$L] | Particles | Particles plastic (%) | Particles organic (%) |
|---|---|---|---|---|---|---|---|
| Nakagusku 2 | S2 | Sth2 | E2 | 40 | 44 | 25.0 | 56.8 |
| Kin 2 | S4 | Sth4 | E4 | 60 | 56 | 17.9 | 66.1 |
| Cape Hedo 1 | S6 | N2 | E6 | 60 | 60 | 13.3 | 66.7 |
| Cape Hedo 2 | S7 | N3 | W1 | 60 | 13 | 15.4 | 61.5 |
| Nago 1 | S8 | N4 | W2 | 60 | 58 | 13.8 | 65.5 |
| Naha 1 | S10 | Sth5 | Naha1 | 40 | 51 | 19.6 | 58.8 |

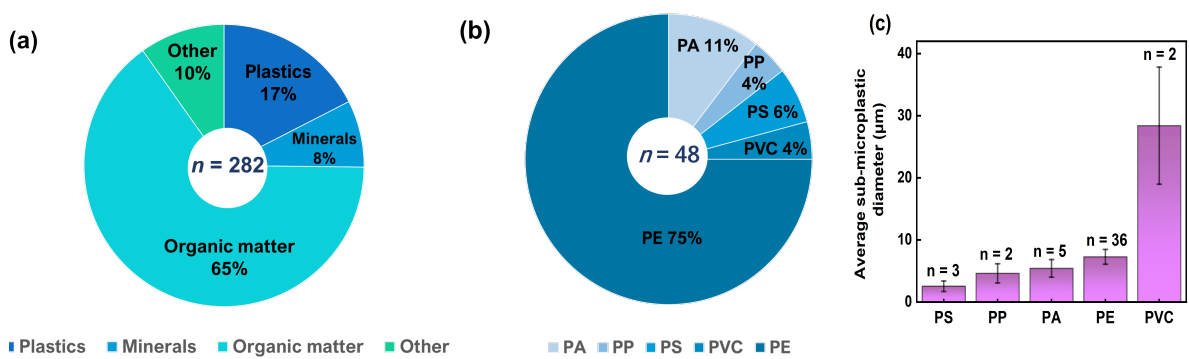

**Figure 5:** (a) Percentage composition of all particles. (b) Polymer types of sub-micron plastics collected in seawater around the main Okinawa island. (c) Average diameter of sub-micron plastics where *n* indicates the number of microplastic polymers.

In the central part of the island, the land use shifts from urban areas to more forest cover [45, 46]. Traffic density goes down by between one third to about one half that found in the southern part of the island [44]. On the east side, we have a station at Kin (S4), while Nago (S8) is located on the west side. Kin has a surrounding population of up to 1,386 people/km$^2$ while at Nago the population density is lower at 296 people/km$^2$. The anthropogenic pressure on the Kin station is predicted to be high [53]. This difference in the population density is reflected in the plastic distribution,

17.9% (S4) and 13.8% (S8), respectively, while the percentage of organic particles is reasonably stable (66.1% (S4) and 65.5% (S8)). At Nago, we analysed 58 particles. PE was the only plastic type found there. In Kin (S4) we characterised a similar number of particles ($n = 56$) but found a wider variety of plastic polymer types (8 PE, one PS, and one PP sub-micron plastics). The particles in Nago bay ranged in size from 1.4 $\mu$m to 27.2 $\mu$m. According to industrial density, which is higher on the east side of the island, the split of the stations into east (S2, S4, S6) and west (S10, S8, S7) does not yield a significant difference (t-test, $p = 0.7$) in sub-micron plastic distribution, as most plastic is correlated with domestic activity, not industrial, on Okinawa.

Finally, in the north of Okinawa, Cape Hedo is located (Figure 1), with a low anthropogenic pressure prediction [53]. We collected particles from two stations (S6 and S7) located on both sides of the cape. In total, 60 particles were identified at station S6, of which 8 are plastics (3 PE, 4 polyamide (PA) and 1 PS) while at station S7 we identified 13 particles with two of them being plastics (1 PE and 1 PA). S6 is located on the east side of the cape, which has rivers draining into the ocean. Because of that, particulate organic matter content is comparable to the station located further south [58]. Polyamide is a family of polymers named Nylon. It is a ductile and strong polymer, permitting the fabrication of textile fibers and cordage. Based on Table 1, PA is the second most common plastic identified in the seawaters of Okinawa with 1.52% of all particles analysed.

Our analysis confirmed that 17% of particles were identified as plastic around Okinawa island, in which PE, PP, PVC, PA and PS are among the most abundant polymer types in aquatic environments (Figure 5(a) and (b)). Polyethylene (PE) was the most common plastic type, comprising of 75% of all the sub-micron plastics polymers analysed (Figure 5(b)). The order of numerical dominance of sub-micron plastic polymers was PE > PA > PS > PP = PVC. Generally, these polymers accounted for 74% of global plastic production and are commonly used in short life-cycle products [42]. Moreover, factors such as hydraulic conditions, salinity, temperature, wind, bio-flouring, as well as changes in surface to volume ratio may affect the distribution of sub-micron plastics around Okinawa.

The source of sub-micron plastics is related to the anthropogenic activities on the seawater, beaches, and in the trading centres in the area around Okinawa. In the fishing communities at the fish landing beaches, woven polymer sacks are used for storage and transport of a variety of products including fishes. Over 75% of the sub-micron plastics are made of polyethylene and these may originate from broken fishing nets, lines or ropes, water bottle caps, household utensils, consumer carry bags, containers/packaging, etc. Recently, a study of the abundance of microplastics in road dust samples collected from several areas in Okinawa shows a high concentration of them in urban areas in which daily vehicle traffic, industrial activity, and high population density are dominant [44]. In the road dust of Okinawa, PE was 29% of the total microplastics [44], while in seawater it is 75% of the total sub-micron plastics. At the end, some of the road dust may be found in the oceans surrounding Okinawa, correlating the two findings via common high concentration areas.

Sub-micron plastics were also classified based on their size as products of degradation of large plastic materials (optical images of each figure). The average size of all collected sub-micron plastics is shown in Figure 5(c). The majority of sub-micron plastics range from 1.4 $\mu m$ to 18.7 $\mu m$, although we identified three microplastics with sizes of 27.2, 30.5, and 47.8 $\mu m$. The smallest average size of $2.53 \pm 0.85$ $\mu m$ is identified for PS polymers while the largest average size of $28.4 \pm 9.4$ is identified for PVC polymers. Likewise, the sampled sub-micron plastics showed a wide range of sizes in various areas of Okinawa with the highest around Naha (S10).

## 4. Summary

In recent years, much progress has been made in understanding the sources, transport, fate, and biological implications of the smallest plastic pollution particles. The public interest in plastic marine pollution and their ecological impacts have increased during the same time. Our results contribute to the knowledge about $in-situ$ analysis and identification of microplastics and demonstrate that the seawater around Okinawa is polluted with micro and sub-micron plastics. They were ubiquitously detected at all sites we tested, with the higher concentration in areas of the island characterised by human activities. All the sub-micron plastics were fragments of plastic materials used by the community, with the major polymers being polyethylene and polyamide materials. While some particles may have originated from and been transported over large distances, correlation with population densities points to land-based sources of the plastic particles. Being predominately found embedded into organic matter, the resulting interactions between marine planktonic organisms and the plastic particles are inevitable. One potential fate could be eventual sedimentation with the rest of the organic matter particle. Concluding, the risks that microplastics pose to fish and their natural foods especially invertebrates, and the possible link to human health, need to be better understood. Strategies such as proper waste management, plastic recycling, and penalties for illegal dumping in areas close to water resources should be promoted and implemented in the communities, to reduce the land-based microplastics found in coastal waters.

## 5. Acknowledgements

This work was funded by Okinawa Institute of Science and Technology Graduate University. DGK acknowledges support from JSPS Grant-in-Aid for Scientific Research (C) Grant Number GD1675001. We thank Elissaios Stavrou from Lawrence Livermore National Laboratory (USA) for useful comments on the Raman peak analysis. We also thank the captain and crew of the Okinawa Prefectural Fisheries and Ocean Research Center ship, Tonan Maru, during the cruise No.672 (FY2018 No.16) and the on-board team of Akinori Murata and Marine Works Japan LTD technician, Shinsuke Toyoda. The Engineering Support Section of OIST is also acknowledged.